\documentclass{ws-procs975x65}

\begin{document}

\title{DEFINITION OF GOOD TETRADS FOR $f(T)$ GRAVITY}

\author{N. TAMANINI$^1$ and C. G. B\"OHMER$^2$}

\address{Department of Mathematics, University College London,\\
London, WC1E 6BT, United Kingdom\\
$^1$E-mail: n.tamanini.11@ucl.ac.uk\\
$^2$E-mail: c.boehmer@ucl.ac.uk}

\begin{abstract}
The importance of choosing suitable tetrads for the study of exact solutions in $f(T)$ gravity is discussed.
For any given metric, we define the concept of good tetrads as the tetrads satisfying the field equations without constrainig the function $f(T)$.
Employing local Lorentz transformations, good tetrads in the context of spherical symmetry are found for Schwarzschild-de Sitter solutions.
\end{abstract}


\bodymatter

\section{Introduction}

In gravitational physics the metric tensor is of paramount importance since it permits to locally measure distances and angles, implying that any experimental result will depend on it. Since the formulation of General Relativity (GR), the majority of gravitational theories assume the metric $g_{\mu\nu}$ to encapsulate the dynamics of the gravitational field. However an equivalent description is provided by tetrad fields $e^a_\mu$ connected to the metric via $g_{\mu\nu}=e^a_\mu e_\nu^b\eta_{ab}$. Here Latin indices refer to the local Lorentz (tangent) space, while Greek indices refer to the spacetime itself (the manifold). Any tedrad $e_\mu^a$ represents four Lorentz vectors which locally connect the tangent space to the curved spacetime.

In gravitational theories which are preserved under both local Lorentz and diffeomorphism transformations, the tetrad and metric formulations are equivalent. However if a theory is not local Lorentz invariant, different Lorentzian observers at the same spacetime point will measure different outcomings from the same experiment. This is the case of $f(T)$ theories of gravity.

This kind of modified gravitational theories is based on the so-called teleparallel equivalent of GR (TEGR), which consists in an equivalent description of GR employing tetrads and torsion. The dynamics is determined by the torsion scalar $T$ which depends on the torsion tensor $T^{\sigma}{}_{\mu\nu} = e_a{}^{\sigma} (\partial_\mu e^a{}_{\nu}-\partial_\nu e^a{}_{\mu})$ and, multiplied by the determinant of the tetrad, forms the Lagrangian of the theory. The Lagrangian of TEGR is equivalent, up to a divergence term, to the GR Lagrangian and thus the two theories are classically equivalent. Moreover $T$ is local Lorentz invariant only up to a divergence, which is enough to ensure that the TEGR action is invariant under local Lorentz transformations.

The situation changes in the case of $f(T)$ gravity which represents theories where the Lagrangian depends on a general function $f$ of $T$. If we now perform a local Lorentz transformation the divergence coming from $T$ cannot become a surface term because of the possible non-linearity of $f$. This implies that $f(T)$ gravity is a non-local Lorentz invariant theory \cite{Li:2010cg}. In particular the gravitational field equations change under local Lorentz transformations while the metric tensor remains unchanged.

\section{Definition of a Good Tetrad}

The situation of $f(T)$ gravity, and in general of any non-local Lorentz invariant theory based on a tetrad formulation, suggests the following question: given a specific metric tensor $g_{\mu\nu}$, does exist a tetrad $e^a_\mu$, solution of the $f(T)$ field equations, such that the relations $g_{\mu\nu}=e^a_\mu e_\nu^b\eta_{ab}$ holds? This question gives rise to the following definition\cite{Tamanini:2012hg}.\\

{\bf Definition.} Given a metric tensor $g_{\mu\nu}$, a {\it good tedrad} $e_\mu^a$ is a tetrad such that
\begin{itemize}
\item $g_{\mu\nu} = e_{\mu}^ae_\nu^b\eta_{ab}$ holds;
\item $e_\mu^a$ is a solution of the $f(T)$ field equations;
\item The function $f(T)$ can be choosen arbitrarily.
\end{itemize}
The last condition is needed for complete generality. It requires that the field equations must be satisfied for an arbitrary function $f(T)$ and not only for specific models.
A good tetrad selects the class of local Lorentz observers who use the metric $g_{\mu\nu}$ to measure physical distances. The concept is of main importance when one deals with exact solutions of the theory.

\section{Schwarzschild-de Sitter Solutions}

As an example we will present a good tetrad associated with the Schwarzschild-de Sitter (SdS) metric
\begin{align}
  ds^2=e^{A(r)}dt^2 -e^{-A(r)}dr^2-r^2d\Omega^2 \quad\mbox{with}\quad e^{A(r)}=\left(1- \frac{2\,M}{r}-\frac{\Lambda}{3}r^2\right) \,,
\label{001}
\end{align}
where $d\Omega^2=d\theta^2+\sin^2\theta d\phi^2$. There is an infinite number of possibile tetrads which give back metric (\ref{001}), namely all the tetrads connected by a local Lorentz transformation. We can thus look for a good tetrad adopting the following procedure\cite{Tamanini:2012hg,Ferraro:2011ks}: first we find a tetrad satisfying $g_{\mu\nu} = e_{\mu}^ae_\nu^b\eta_{ab}$ and then we use a local Lorentz transformation in order to ensure that also the other requirements are satisfied.

The simplest possible tetrad related to the SdS solution is of course the diagonal one
\begin{align}
e_\mu^a = \mbox{diag}(e^{A/2},e^{-A/2},r,r\sin\theta) \,,
\label{002}
\end{align}
where $A$ is given in (\ref{001}). Unfortunately this tetrad is not a solution of the $f(T)$ field equations and even for other spherically symmetric metrics it represents a solution only in the $f''(T)=0$ case\cite{Boehmer:2011gw} which is against the third point of our definition.

Even if (\ref{002}) is not a good tetrad we can still apply a local Lorentz transformation to find a good one. We will concentrate on local rotations given by the transformation matrix
\begin{align}
  {\Lambda^a}_b=
  \begin{pmatrix}
    1 & 0 \\
    0 & \mathcal{R}(\varphi,\vartheta,\psi) \\
  \end{pmatrix}\,,
  \label{003}
\end{align}
where $\mathcal{R}(\varphi,\vartheta,\psi)$ is a three-diensional rotation characterized by the three Euler angles $\varphi$, $\vartheta$, $\psi$ which are general functions of the spacetime coordinates. For our scope it is enough to consider only the following values
\begin{align}
  \varphi = \gamma(r) \,,\quad
  \vartheta = \theta-\pi/2\,, \quad 
  \psi = \phi\,.
\end{align}
Then under the transformation (\ref{003}) the diagonal tetrad (\ref{002}) changes into\cite{Tamanini:2012hg}
\begin{multline}
\hat e_\mu{}^a= \Lambda_b{}^a{e_\mu}^b = 
\left(
\begin{array}{cc}
 e^{A/2} & 0 \\
 0 & e^{-A/2} \sin \theta \cos \phi  \\
 0 & -r \left(\cos \theta \cos \phi \sin \gamma+\sin \phi \cos\gamma \right) \\ 
 0 & r \sin \theta \left(\sin \phi \sin\gamma -\cos \theta \cos \phi \cos\gamma \right)
\end{array}\right.\\
\left.
\begin{array}{cc}
 0 & 0 \\
 e^{-A/2} \sin \theta \sin \phi  & e^{-A/2} \cos \theta  \\
r \left(\cos \phi \cos\gamma -\cos \theta \sin \phi \sin\gamma \right) & r
   \sin \theta \sin\gamma  \\
 -r \sin \theta \left(\cos \theta \sin \phi \cos\gamma +\cos
   \phi \sin\gamma \right) & r \sin ^2\theta \cos\gamma 
\end{array}
\right) \,.
\label{006}
\end{multline}
In $f(T)$ gravity the particular case $\gamma = -\pi/2$ was first considered for spherically symmetric spacetimes in Ref.~\refcite{Boehmer:2011gw}, where however was noticed that it is not a solution of the field equations, though it helps to relax the $f''(T)=0$ constraint arising with the diagonal tetrad. A good tetrad for the SdS metric can be found from (\ref{006}) if one chooses the function $\gamma$ such that the torsion scalar becomes constant\cite{Tamanini:2012hg}. In this case the tetrad (\ref{006}) is a solution of the field equations provided $T=T_0$ satisfies
\begin{align}
  \Lambda = \frac{1}{2} \left(\frac{f(T_0)}{f'(T_0)}-T_0\right) \,,
\label{004}
\end{align}
with $T_0$ 
always gauged by the function $\gamma$. In other words for any value of $\Lambda$ one can always find a $\gamma$ such that (\ref{004}) is true. Note that in the TEGR case we have $\Lambda=0$ recovering the Schwarzschild solution as required by the equivalence with GR.

\section{Conclusions}

The definition of good tetrads for $f(T)$ gravity is a useful tool when one deals with exact solutions of the theory. Given a particular metric, it permits to define the class of local Lorentz observers who use that metric to measure distances. In non-local Lorentz invariant theories, such as $f(T)$ gravity, different Lorentz observers will not agree on the outcomes of gravitational experiments. It is important thus to determine the observers who employ physical metrics, such as the SdS metric, for measurements. A good tetrad defines such observers in $f(T)$ gravity.

\end{document}